\documentclass[sigconf]{acmart}
\AtBeginDocument{%
  \providecommand\BibTeX{{%
    \normalfont B\kern-0.5em{\scshape i\kern-0.25em b}\kern-0.8em\TeX}}}

\usepackage{float}
\usepackage[most]{tcolorbox}

\newfloat{prompt}{htbp}{locf}
\floatname{prompt}{Prompt}
\newcounter{promptcounter}
\newtcolorbox[use counter=promptcounter]{promptbox}[2][]{
  enhanced, breakable, floatplacement=htb, title={\textbf{Prompt {\thepromptcounter}: #2}},
  width=\columnwidth, top=1mm,left=1mm,  right=1mm, bottom=1mm, boxrule=1pt, arc=1mm,
  fontupper=\small,
  #1
}
\usepackage{textcomp}
\usepackage{listings}
\usepackage{xcolor}

\definecolor{backcolour}{rgb}{0.95,0.95,0.92}

\lstdefinestyle{mystyle}{backgroundcolor=\color{backcolour}, basicstyle=\ttfamily\footnotesize,
    breakatwhitespace=false, breaklines=true, captionpos=b, keepspaces=true,                 
    numbers=left, numbersep=5pt, showspaces=false, showstringspaces=false,
    showtabs=false, tabsize=1}

\setcopyright{acmlicensed}
\copyrightyear{2024}
\acmYear{2024}
\acmDOI{XXXXXXX.XXXXXXX}

\acmConference[ASE'24]{39th IEEE/ACM International Conference on Automated Software Engineering}{October 27 - November 1, 2024}{Sacramento, California, USA.}

\begin{document}

\title{Getting Inspiration for Feature Elicitation: \\ App Store- vs. LLM-based Approach}

\author{Jialiang Wei}
\email{jialiang.wei@mines-ales.fr}
\orcid{0009-0008-6028-1576}
\affiliation{
  \institution{EuroMov Digital Health in Motion, Univ Montpellier, IMT Mines Ales}
  \city{Ales}
  \country{France}
}

\author{Anne-Lise Courbis}
\email{anne-lise.courbis@mines-ales.fr}
\orcid{0000-0002-7530-4661}
\affiliation{
  \institution{EuroMov Digital Health in Motion, Univ Montpellier, IMT Mines Ales}
  \city{Ales}
  \country{France}
}

\author{Thomas Lambolais}
\email{thomas.lambolais@mines-ales.fr}
\orcid{0000-0002-4462-1277}
\affiliation{
  \institution{EuroMov Digital Health in Motion, Univ Montpellier, IMT Mines Ales}
  \city{Ales}
  \country{France}
}

\author{Binbin Xu}
\email{binbin.xu@mines-ales.fr}
\orcid{0000-0002-0822-5250}
\affiliation{
  \institution{EuroMov Digital Health in Motion, Univ Montpellier, IMT Mines Ales}
  \city{Ales}
  \country{France}
}

\author{Pierre Louis Bernard}
\email{pierre-louis.bernard@umontpellier.fr}
\orcid{0000-0002-0023-4531}
\affiliation{
  \institution{EuroMov Digital Health in Motion, Univ Montpellier, IMT Mines Ales}
  \city{Montpellier}
  \country{France}
}

\author{Gérard Dray}
\email{gerard.dray@mines-ales.fr}
\orcid{0000-0003-1525-5682}
\affiliation{
  \institution{EuroMov Digital Health in Motion, Univ Montpellier, IMT Mines Ales}
  \city{Ales}
  \country{France}
}

\author{Walid Maalej}
\email{walid.maalej@uni-hamburg.de}
\orcid{0000-0002-9550-0037}
\affiliation{
  \institution{University of Hamburg}
  \city{Hamburg}
  \country{Germany}
}

\renewcommand{\shortauthors}{Wei et al.}

\begin{abstract}
Over the past decade, app store (AppStore)-inspired requirements elicitation has proven to be highly beneficial.
Developers often explore competitors' apps to gather inspiration for new features.
With the advance of Generative AI, recent studies have demonstrated the potential of large language model (LLM)-inspired requirements elicitation.
LLMs can assist in this process by providing inspiration for new feature ideas.
While both approaches are gaining popularity in practice, there is a lack of insight into their differences. 
We report on a comparative study between AppStore- and LLM-based approaches for refining features into sub-features. 
By manually analyzing 1,200 sub-features recommended from both approaches, we identified their benefits, challenges, and key differences.
While both approaches recommend highly relevant sub-features with clear descriptions, LLMs seem more powerful particularly concerning novel unseen app scopes. 
Moreover, some recommended features are imaginary with unclear feasibility, which suggests the importance of a human-analyst in the elicitation loop.
\end{abstract}



\maketitle

\section{Introduction}
Requirements Elicitation (or Requirements Development) aims to identify and understand the requirements of a system and the needs of its stakeholders \cite{Maalej:ManagingRequirementsKnowledge:2013}. 
This process can be implemented using various techniques such as interviews, questionnaires, and observations \cite{Dieste:SystematicReviewAggregation:2011}.
Within this process, feature elicitation specifically refers to gathering information to identify and define the system features that should be implemented to fulfill the stakeholders' needs.

Over the past decade, the popularity of mobile devices has led to a substantial increase in the number of apps available on various app stores.
According to Statista, until 2023, Apple's App Store offers 4.83 million apps\footnote{\url{https://www.statista.com/statistics/268251/number-of-apps-in-the-itunes-app-store-since-2008/}}, while Google Play has 2.43 million apps\footnote{\url{https://www.statista.com/statistics/266210/number-of-available-applications-in-the-google-play-store/}}.
App stores include valuable data that can serve as a source of inspiration for feature elicitation \cite{Ferrari:StrategiesBenefitsChallenges:2023, Martens:ReleaseEarlyRelease:2019, Johann:SAFESimpleApproach:2017}. 

Consider, for example, the following scenario: 
Jay, a mobile app developer, wishes to create a new app  for sleep tracking. 
His initial concept of the app may be quite rudimentary. 
The envisioned app might require integration with a smartwatch to monitor physiological metrics such as heart rate, breathing patterns, and other related parameters. 
However, transforming these initial ideas into concrete app features necessitates a cycle of refinement and development.
One practical approach for Jay is to examine existing apps in the same domain to identify features that have already been implemented successfully. 
Recent empirical evidence underscores the prevalence of this strategy. 
A study by Jiang et al.~found that 86.1\% of app developers take into account the features of similar apps when developing their own apps \cite{Jiang:RecommendingNewFeatures:2019}.
Over the past decade, research has suggested various approaches for AppStore-inspired feature elicitation.
Numerous studies have focused on feature recommendation \cite{Liu:MethodAcquireCrossdomain:2022,Wang:MissingStandardFeatures:2022,Jiang:RecommendingNewFeatures:2019,Chen:RecommendingSoftwareFeatures:2019,Liu:InformationRecommendationBased:2019} or competitive analysis \cite{Assi:FeatCompareFeatureComparison:2021,Uddin:AppCompetitionMatters:2020,Dalpiaz:RESWOTUserFeedback:2019,Shah:UsingAppReviews:2019} by mining app descriptions, app reviews, and the user interfaces.

Recently, researchers also started using Large Language Models (LLMs) for getting inspirations and recommendations for requirements \cite{Arora:AdvancingRequirementsEngineering:2023}.
These advanced models, trained on internet-scale knowledge, enable the automated generation of text, which can be leveraged to create user stories, goal models, and other requirements artifacts.
Researchers used LLMs, e.g.~ChatGPT, 
to generate user stories that describe candidate human values providing inspiration to stakeholder discussions  \cite{Marczak-Czajka:UsingChatGPTGenerate:2023} or to refine  user stories and improve their quality \cite{Zhang:LLMbasedAgentsAutomating:2024}.
Others explored the capacity of ChatGPT on generating goal models from a given context description \cite{Chen:UseGPT4Creating:2023,Nakagawa:MAPEKLoopbasedGoal:2023}.
Overall, LLMs seem capable of boosting efficiency and creativity in requirements elicitation. 
In fact, this empirical study demonstrates the significant capacity of LLMs for feature elicitation.

While the use of both app stores and LLMs demonstrated promising results for generating new features, there is a limited insight into their effectiveness. 
This study aims to fill this gap by examining the benefits, challenges, and differences between these two approaches. 
We focus on feature refinement, a specific task of feature elicitation, which involves breaking down a high-level feature, such as ``health monitoring'' into a list of lower level sub-features like ``sleep tracking'', ``heart rate monitoring'', and ``nutrition logging''.
In the LLM-based approach, sub-features are generated directly by prompting GPT-4. 
The AppStore-based approach involves searching for relevant app descriptions in a vector database, extracting features from these descriptions, and then selecting relevant features. 
To ensure a fair comparison, we also use GPT-4 for the extraction and selection steps in the AppStore approach.

To compare the two approaches, we studied 20 high-level root features, including 10 already existing features and 10 novel features that have not been implemented elsewhere. 
For each of these 20 root features, we automatically generated two two-levels feature trees: one LLM-based and the other AppStore-based. 
This resulted in a total of 40 feature trees, including 1,200 sub-features. 
Each of the 1,200 sub-features was then manually assessed for relation to its super feature, relevance, clarity, traceability, and feasibility.
Further, we evaluated the intersection sets and the difference sets of the features generated from the two approaches.

This paper's contribution is threefold. 
First, it presents a detailed comparison between LLM-inspired and AppStore-inspired feature elicitation. 
Second, it provides insights on how to effectively utilize both approaches. 
Third, it introduces a tool that integrates both approaches\footnote{\url{https://github.com/Jl-wei/feature-inspiration}}.
In the following we introduce the LLM- and AppStore-based approaches with corresponding prompts in Section \ref{sec:llm} and Section \ref{sec:appstore}. 
Then, we present our study design in Section \ref{sec:design} and report on the results in Section \ref{sec:results}.
Finally, Section \ref{sec:disc} discusses the findings, tool support, and the threats to validity while Section \ref{sec:relwork} summarizes related work and Section \ref{sec:conclusion} concludes the paper.

\begin{figure}[!t]
\centering
\includegraphics[width=0.9\columnwidth]{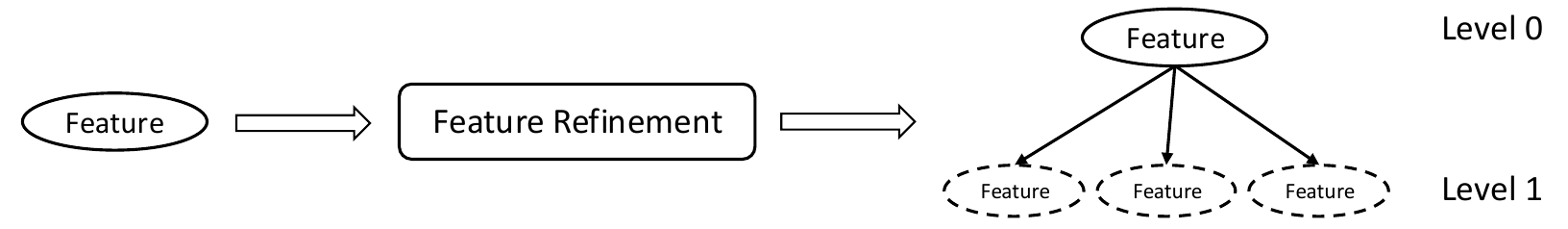}
\caption{Illustration of a single feature refinement.}
\Description{}
\label{fig:single-input}
\end{figure}

\begin{figure}[!t]
\centering
\includegraphics[width=1\columnwidth]{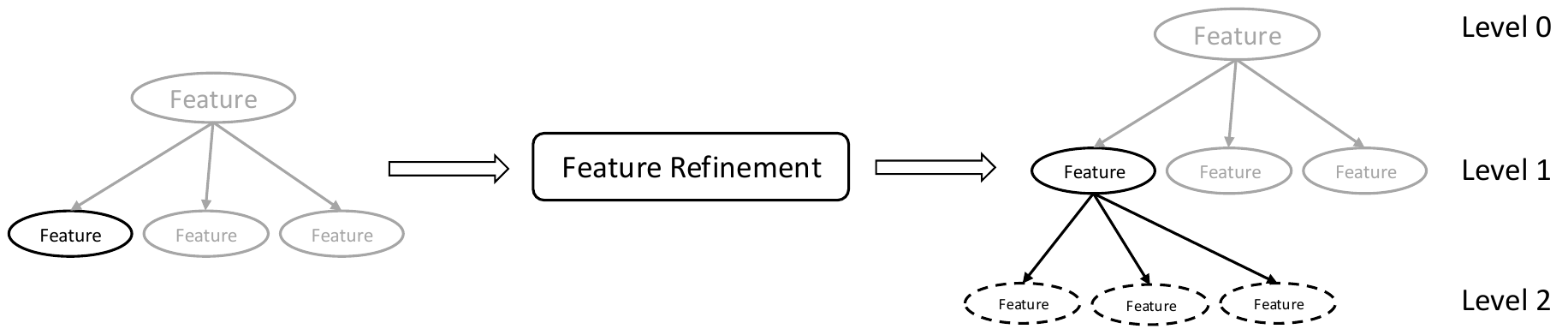}
\caption{Illustration of feature refinement with its context (i.e.~its super feature and sibling features).}
\Description{}
\label{fig:context-input}
\end{figure}

\begin{figure*}[]
\centering
\includegraphics[width=\textwidth]{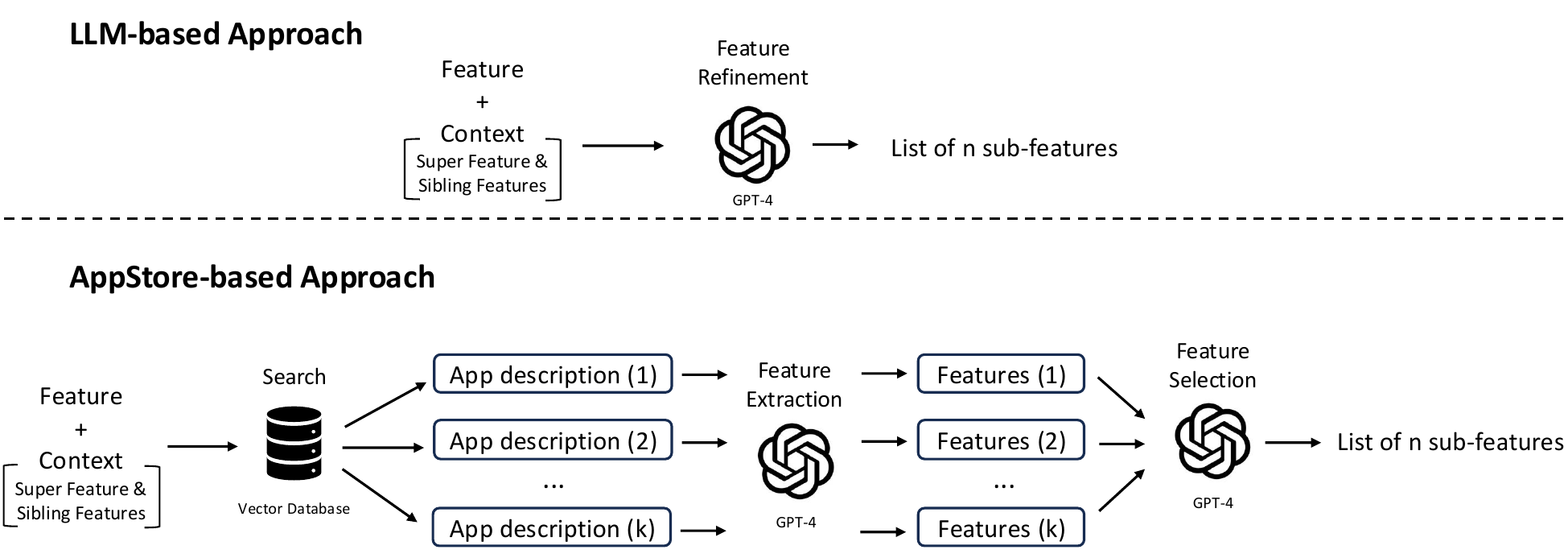}
\caption{LLM-inspired vs.~AppStore-inspired feature refinement (the context of a feature is its super feature + sibling features).}
\Description{}
\label{fig:overview}
\end{figure*}

\section{LLM-Based Inspiration}
\label{sec:llm}
Large language models are AI models designed to understand, generate, and manipulate human language. 
They are pre-trained for ``next-word prediction'' (next-token) on vast amounts of text data from diverse sources, such as books, articles, websites, and other text repositories, which allows them to understand a wide range of topics \cite{Zhao:SurveyLargeLanguage:2023}.
Our implementation of LLM-based approach is accomplished by prompting GPT-4 \cite{OpenAI:GPT4TechnicalReport:2024}, which is at the time of this research one of the most advanced LLMs.
As shown on Figure \ref{fig:overview}, the model takes a feature as input and refines it to a list of sub-features.

To enhance the model performance, Prompt \ref{prompt:system-llm} is employed as the system prompt, thereby assigning a specific role to GPT-4\footnote{\url{https://platform.openai.com/docs/guides/prompt-engineering/tactic-ask-the-model-to-adopt-a-persona}}.
\begin{promptbox}[label=prompt:system-llm]{System prompt}
You are an expert in mobile app development and requirements engineering. 
You excel at decomposing high-level features into detailed sub-features.
\end{promptbox}

There are two scenarios for feature refinement. 
In the first scenario, a single feature and its description are provided as input, as illustrated in Figure \ref{fig:single-input}.
In this scenario, the approach recommends sub-features based on the information provided about this feature.
Prompt \ref{prompt:inspiration-feature} takes the feature and its description as input, allowing the model to generate \textit{n} corresponding sub-features that are formatted as a JSON list to facilitate further processing.

\begin{promptbox}[label=prompt:inspiration-feature]{LLM refinement of a single feature}
**Feature**

\textasciigrave\textasciigrave\textasciigrave

\{feature\}: \{feature\_description\}

\textasciigrave\textasciigrave\textasciigrave

Given the mobile app feature above, please refine it to a list of sub-features.

Ensure that the number of sub-features is \{n\}.

The output should be a list of JSON formatted objects like this:

[\{
"sub-feature": sub-feature,
"description": description
\}]
\end{promptbox}

The previous scenario focuses only on the refinement of one specific feature without considering its broader context.
This can be appropriate for refining a root feature. 
In the second scenario, both the feature and its super feature and sibling features are provided as input, as shown on Figure \ref{fig:context-input}.
The \textit{super} feature refers to the overarching feature that encompasses the specific feature in question, while the \textit{sibling} features are those that exist at the same hierarchical level as the specific feature.
By including these additional elements as shown in Prompt \ref{prompt:inspiration-superfeature}, GPT-4 should recommend sub-features that are not only relevant to the root feature but also harmonious with the overall app design.

\begin{promptbox}[label=prompt:inspiration-superfeature]{LLM refinement of a feature with its context (i.e.~its super feature and sibling features)}
**Super Feature**

\textasciigrave\textasciigrave\textasciigrave

super-feature: \{super\_feature\}

description: \{super\_feature\_description\}

\textasciigrave\textasciigrave\textasciigrave

Knowing that the feature "\{super\_feature\}" above is refined into a list of the following features:

\textasciigrave\textasciigrave\textasciigrave

\{sub\_features\}

\textasciigrave\textasciigrave\textasciigrave

Please refine the following feature to a list of sub-features.

Ensure that the number of sub-features is \{n\}.

**Feature**

\textasciigrave\textasciigrave\textasciigrave

\{feature\_with\_desc\}

\textasciigrave\textasciigrave\textasciigrave

The output should be a list of JSON formatted objects like this:

[\{
"sub-feature": sub-feature,
"description": description
\}]
\end{promptbox}

\section{AppStore-Based Inspiration}
\label{sec:appstore}
As illustrated on Figure \ref{fig:overview}, the AppStore-inspired feature refinement includes three steps:
(1) search for relevant descriptions on an app description repository, (2) extract pertinent app features from these app descriptions, 
and (3) select sub-features from the extracted features.

\subsection{Searching the App Descriptions}
In this study, instead of relying on the Google Play search engine to find relevant app descriptions, we developed a custom app description search engine. 
The Google Play search engine, as our tests indicate, suffers from two main issues.
First, it often struggles with complex or lengthy queries, frequently returning completely irrelevant app descriptions.
Second, the search results are inconsistent and not reproducible, varying with each search attempt. 
According to the Google Play documentation\footnote{\url{https://support.google.com/googleplay/android-developer/answer/9958766?hl=en&sjid=4123634560946816541-EU}}, the ranking of search results may be influenced by factors such as user relevance, app quality, editorial value, and advertisements as well.
However, our objective is to acquire the most semantically relevant app descriptions relative to the query.

\begin{figure}[!t]
\centerline{\includegraphics[width=0.5\textwidth]{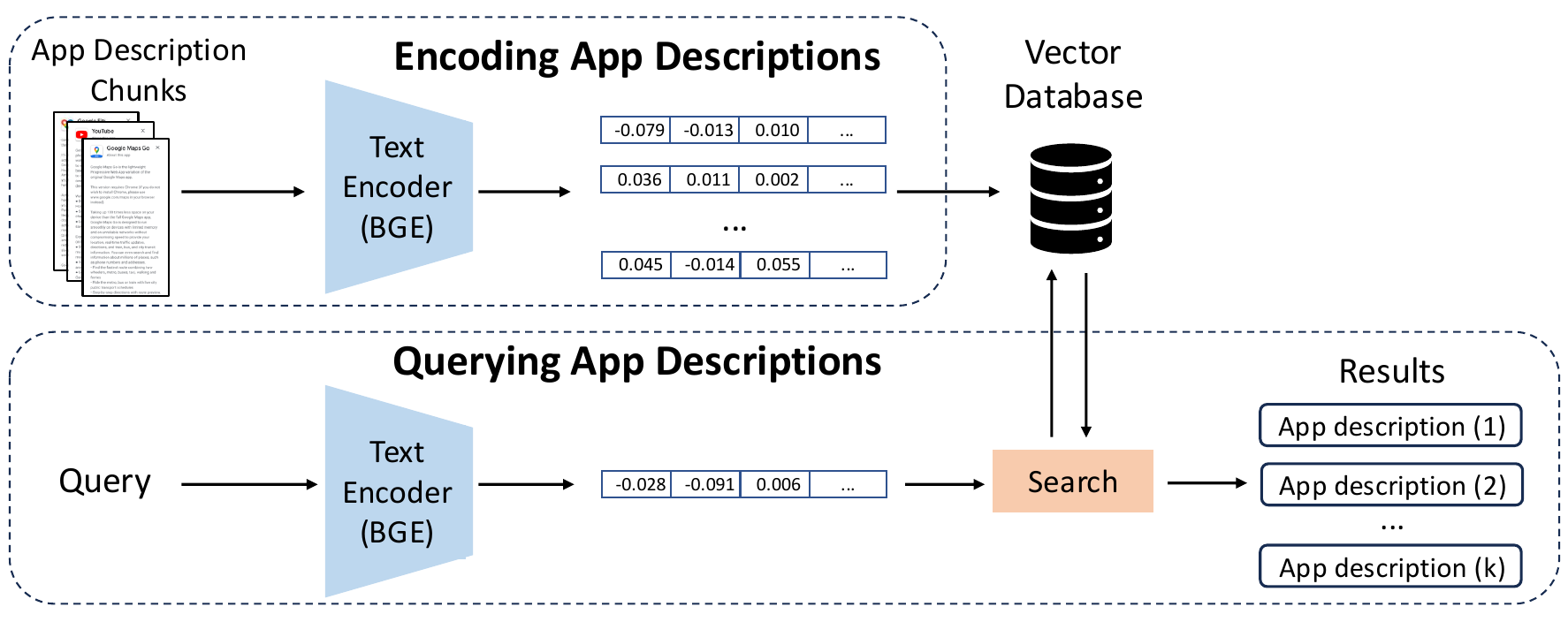}}
\caption{Encoding and querying the app descriptions.}
\Description{}
\label{fig:search}
\end{figure}

Our custom search engine is designed to address these shortcomings by focusing specifically on semantic relevance, thereby ensuring that the retrieved app descriptions are closely aligned with the query.
To develop our own search engine, we collected a comprehensive repository of app descriptions. 
These descriptions were encoded into text embeddings and stored in a vector database, enabling efficient querying, as shown on Figure \ref{fig:search}.

\subsubsection{App Description Collection}
Given the ID of an app, one can easily get its description with Google Play Scraper\footnote{\url{https://github.com/facundoolano/google-play-scraper}}.
Since Google Play does not provide a comprehensive list of all available apps, we developed a strategy to collect as many app IDs as possible.
Our data collection strategy is divided into two steps:
\begin{enumerate}
\item We conducted searches on Google Play using each word in an English dictionary\footnote{\url{https://github.com/mwiens91/english-words-py}} as the query.
The dictionary comprises 114,769 words, resulting in 114,769 searches on Google Play \footnote{To mitigate load pressure on Google Play's servers, we made one query per minute.}.
Each search yields a maximum of 30 apps.
\item The apps collected in the first step served as a seed list. 
For each app, Google Play often provides recommendations for similar apps and lists other apps developed by the same developer. 
This scenario can be conceptualized as a graph, where apps are represented as nodes and the relationships (such as app similarity and common developer) are represented as edges. 
Starting from the seed list, we performed a breadth-first search of this graph.
\end{enumerate}
We finally collected a total of 849,260 distinct apps (as of Feb.~2024).

\subsubsection{App Description Filtering}
To ensure the suitability of app descriptions for our analysis, we employed a filtering process:

\begin{itemize}
\item \textit{Remove games}:
We focus our work on feature elicitation for regular apps. 
Previous work suggest that game descriptions tend to be different \cite{Guzman:HowUsersThis:2014}. 
We excluded those to prevent potential bias in the results.

\item \textit{Remove non-English descriptions}: 
Despite collecting apps from Google Play in the USA, some descriptions may not be in English. 
Since our focus is on English-language, we excluded all non-English entries using Lingua\footnote{\url{https://github.com/pemistahl/lingua-py}}.

\item \textit{Remove too short descriptions}:
App descriptions that are too short do not provide sufficient information for feature extraction. 
Therefore, we removed all app descriptions shorter than 200 characters.
\end{itemize}
After the filtering process, a total of 589,363 apps remained.

\subsubsection{App Descriptions Encoding}
The objective of this process is to convert app descriptions into text embeddings to facilitate semantic search. 
For this purpose, we utilized BGE \cite{Xiao:CPackPackagedResources:2023} as the embedding model. 
BGE, as described by Xiao et al., is a state-of-the-art text embedding model\footnote{\url{https://huggingface.co/spaces/mteb/leaderboard}}. 
It is trained through a three-step process: pre-training with plain text, contrastive learning on a text pair dataset, and task-specific fine-tuning. 
Considering that the maximum input length for BGE is 512 tokens, equivalent to approximately 2000 characters
, we initially divided each app description into chunks with a maximum length of 2000 characters.
Then, each chunk was encoded with BGE into embeddings of 384 dimensions and stored in our vector database.

\subsubsection{App Descriptions Querying}
During the querying phase, the textual query is encoded using BGE, producing a 384-dimensional vector. 
This query embedding is then employed to retrieve the top-k most similar description embeddings from the vector database. 
In our study, we utilized cosine similarity as the metric for assessing similarity. 
Consequently, the query results consist of the top-k descriptions that exhibit the highest degree of resemblance.

In the first scenario, shown in Figure \ref{fig:single-input}, we concatenated the target feature name and its description as the query.
In the second scenario, shown in Figure \ref{fig:context-input}, the query was constructed by concatenating the name and description of both the feature and its super feature separated by a semicolon.

\subsection{Extracting App Features}
To ensure a fair comparison, we used a similar system prompt to that of LLM-based approach, with the following additional sentence specifically for app feature extraction: 
``Additionally, your expertise extends to extracting app features from descriptions, enabling you to identify key functionalities like "step count", "group chats", and "multi-device synchronization".''

The acquired app descriptions are subsequently processed using GPT-4 in a map-reduce manner. 
In the single feature scenario (Figure \ref{fig:single-input}), each app description is examined to extract features pertinent to the query feature through Prompt \ref{prompt:extraction}. 
This prompt accepts an app description along with a feature name and its description as input and returns a list of JSON objects.  
Each JSON object contains the name and description of the sub-features extracted from the app description.
\begin{promptbox}[label=prompt:extraction]{AppStore feature extraction}
**App description**

\textasciigrave\textasciigrave\textasciigrave

\{app\_description\}

\textasciigrave\textasciigrave\textasciigrave

From the app description above, please extract the sub-features of this following feature.

Ensure that all sub-features are from the app description.

**Feature**

\textasciigrave\textasciigrave\textasciigrave

\{feature\_with\_desc\}

\textasciigrave\textasciigrave\textasciigrave

The output should be a list of JSON formatted objects like this:

[\{
"sub-feature": sub-feature,
"description": description
\}]
\end{promptbox}

The resulting JSON objects are subsequently processed to include the "source-app-id" field through a Python script, which indicates the app ID from which the sub-feature was extracted.
For each app description, we obtain a JSON list where each object contains the fields "sub-feature", "description", and "source-app-id". 
The prompt, which takes feature with its super feature and sibling features as input (Figure \ref{fig:context-input}) to extract sub-features, is available in the source code of our proposed tool.

\subsection{Selecting Sub-Features}
From the previous step, we get \textit{k} lists of sub-features from the corresponding \textit{k} app descriptions.
The total number of sub-features may easily exceed 20, which would be excessive for inspiration. 
Additionally, this list often includes duplicates and less relevant items.
To tackle this problem, we applied Prompt \ref{prompt:selection} for feature selection.
It merges the \textit{k} lists of sub-features into one single list and selects the \textit{n} most relevant ones based on their descriptions.
In the end, the result is a JSON list containing \textit{n} sub-features.
\begin{promptbox}[label=prompt:selection]{AppStore feature selection}
\textasciigrave\textasciigrave\textasciigrave json

\{features\}

\textasciigrave\textasciigrave\textasciigrave

Given the JSON lists of app features provided above, please combine them into a single list. 

Ensure that similar sub-features are merged into one.

You should only keep \{n\} sub-features that are most relevant to the following feature description:

\textasciigrave\textasciigrave\textasciigrave

\{feature\_with\_desc\}

\textasciigrave\textasciigrave\textasciigrave

The output should be a list of JSON formatted objects like this:

[\{
"sub-feature": sub-feature,
"description": description,
"source-app-id": source-app-id
\}]
\end{promptbox}

\section{Evaluation Design}
\label{sec:design}
Our evaluation focuses on the following research question:

\textbf{RQ}: How good are the generated features and what are the differences between LLM-based and AppStore-based approaches? 

To answer this question, we prepared 20 app features across various domains (as root features). 
Each root feature was used separately as input for LLM-Inspiration and AppStore-Inspiration to generate two two-levels feature trees. 
Subsequently, three authors \textit{independently} evaluated the quality of \textit{all} generated sub-features.

\subsection{Root Features Preparation}
App developers might aim to implement both existing features from other apps and novel features that have not been previously implemented.
To explore both situations, we selected 10 existing features and devised 10 novel features as presented in Table \ref{tab:root-features}.

\begin{table}[!htp]
\caption{Root features used in the evaluation.}
\small
\begin{tabular}{ll}
\toprule
\textbf{Existing features} & \textbf{Novel features} \\
\midrule
Anti Smartphone Addiction & Contextual Soundscape \\
Criminal Alert & Driver Guardian \\
Interior Design & Interactive Historical Overlay \\
Mental Health Therapy & Laugh evaluation \\
Parking Space Finder & Mood-Adaptive UI \\
Random Chat & Predictive Subscription Management \\
Supermarket Checkout & Social Health Analytics \\
Travel Planner & Symbiotic Music Creation \\
Virtual Fashion Assistant & Synesthetic Sensory Augmentation \\
Voice Translation & Thought reading \\
\bottomrule
\end{tabular}
\label{tab:root-features}
\end{table}

The 10 existing features were chosen from a recent BuildFire article, which proposed 50 interesting app features for 2024\footnote{\url{https://buildfire.com/best-app-ideas/}}. 
From these 50 features, we selected 10 using specific criteria: 
the features should not be too high-level (e.g., education, social networking) or overly popular (e.g., dating, job search, eBook reader). 
Additionally, these features should be sourced from diverse categories, according to the taxonomy provided by Google Play\footnote{\url{https://support.google.com/googleplay/android-developer/answer/9859673}}. 
Then, we created succinct description for the 10 features by summarizing the presentation from the article.
We searched the 10 features on Google Play, and could confirm that each of them has been mentioned in multiple app descriptions.

The 10 novel features were generated through a collaborative brainstorming session. 
In a meeting room, three authors contributed ideas by writing on a whiteboard, drawing from their individual experiences, current trends, and emerging technologies. 
From these ideas, we selected 10 that we agreed were particularly innovative and were from 10 different app categories.
Each idea was then discussed and refined collaboratively into a feature name and a succinct description.
We searched these 10 features on Google Play, and could confirm that no existing apps have mentioned them in their app descriptions.

\begin{figure}[!h]
\centerline{\includegraphics[width=0.4\textwidth]{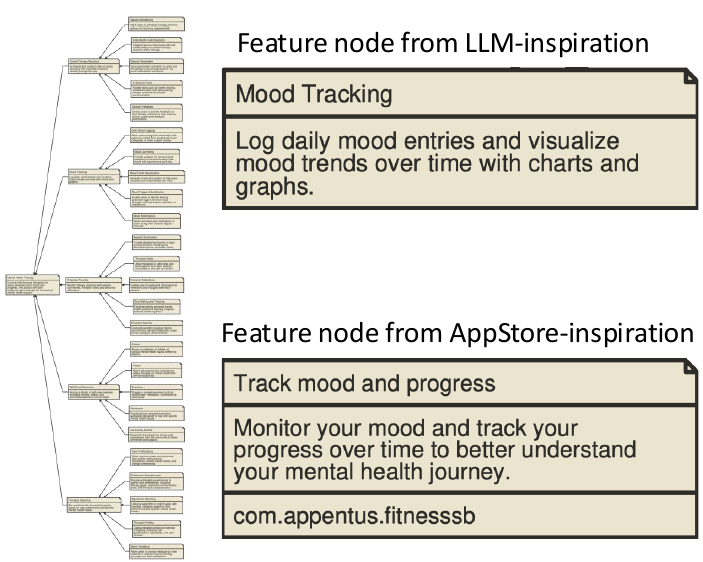}}
\caption{Example of feature tree and feature nodes.}
\Description{}
\label{fig:feature-tree}
\end{figure}

\subsection{Feature Trees Generation}
Once the root features were prepared, we applied the LLM-Inspiration and AppStore-Inspiration approaches to each root feature to generate two-levels feature trees.
The only input for the generation of a feature tree is the root feature along with its description. 
We did not add any additional ``app context'' as input for the tree generation, because the description associated with the root feature represents by itself the app context. 
For example, the description of the root feature ``Travel Planner'' is: ``Plan perfect trip from flights to personalized itineraries with this travel app that offers bookings, reviews, and recommendations for restaurants, attractions, and activities.'' 
All generations of the sub-features were performed automatically without human intervention.
To ensure the generated features are assessable, we set the number of relevant app description \textit{k} to 3 and the number of generated sub-features \textit{n} to 5.
That is, we generated five sub-features for each root feature (L0), and for each sub-feature (L1), five additional (sub-)sub-features (L2). 
Consequently, each generated feature tree contained a total of 30 sub-features. 
Finally, we generated a total of 40 feature trees: 20 utilizing the LLM-Inspiration and 20 employing the AppStore-Inspiration.
Figure \ref{fig:feature-tree} illustrates the feature tree and the feature nodes obtained with both approaches.
The node from AppStore-Inspiration includes an additional field that displays the source app ID of the feature.
The generated trees were then used in the subsequent evaluation.

\subsection{Feature Quality Evaluation}
Three authors manually evaluated the quality of all generated sub-features.
Each author independently evaluated a total of 40 feature trees: comprising 10 existing and 10 novel root features; for which one tree is generate with LLM- and one with AppStore-Inspiration.

\subsubsection{Feature Node Evaluation}
To assess the quality of the generated sub-features, we employed the following evaluation metrics:

\begin{itemize}
\item \textit{Relationship with Super Feature}: 
What is the relationship between the generated sub-feature and its super feature? 
Is the generated sub-feature truly subordinate to the super feature, or is it instead a sibling feature, a parent feature, an identical feature, or other type of relationship?

\item \textit{Relevance}: 
How closely does the sub-feature relate to its root features? 
The relevance metric ensured that the generated sub-features were pertinent and logical extensions of its root feature.

\item \textit{Clarity}:
How well is the sub-feature described? 
The clarity metric assessed how easily developers could understand and interpret the generated sub-feature descriptions.
\end{itemize}

For sub-features obtained with LLM-Inspiration, we also evaluate their feasibility. 
We did not evaluate the feasibility of the sub-features generated with AppStore-Inspiration, as these are sourced from existing apps and are thus inherently technically feasible.
\begin{itemize}
\item \textit{Feasibility}:
Is the sub-feature technically and practically feasible to implement? 
This metric evaluated whether the generated sub-features were realistic from a technical and practical standpoint (to the knowledge of the evaluator).
\end{itemize}

Additionally, for the sub-features obtained using AppStore-Inspi\-ration, we assess their traceability. 
The feature extraction and feature selection of AppStore-Inspiration were performed with the help of GPT-4. 
However, due to the potential hallucination issues associated with GPT \cite{Rawte:SurveyHallucinationLarge:2023}, there is a possibility that some of these features were not derived from the app descriptions but were instead fabricated by the model.
Given the app ID associated with AppStore-Inspired features, we can compare each feature against its original app description to assess its traceability.
\begin{itemize}
\item \textit{Traceability}:
Does the sub-feature originate from the corresponding app description, or is it a fabrication created by the LLM?
This metric evaluated whether the sub-features were extracted from app descriptions.
\end{itemize}

\subsubsection{Feature Tree Evaluation}
In addition to evaluating each feature node, we manually assessed the entire trees, focusing on:
\begin{itemize}
\item \textit{Number of Distinct Features}:
This metric quantifies the number of distinct features within the generated feature tree.
This metric aims to address the issue of duplicated features.

\item \textit{Number of Distinct Relevant Features}:
Similar as the \textit{Number of Distinct Features}, but only features with a \textit{relevance} score of 4 or higher are counted. 

\item \textit{Number of Common Relevant Feature of Both Approaches}:
This metric quantifies the relevant feature that are generated by both approaches.
\end{itemize}

\subsubsection{Evaluation Protocol}
We followed a common content analysis protocol \cite{Krippendorff:ContentAnalysisIntroduction:2018} during our evaluation including three steps. 

First, we met twice to discuss the root features and the evaluation metrics and to create a shared understanding using examples. 
This resulted in an evaluation guideline that defines the metrics (introduced in the previous section) and the semantic scale to assess them as shown on Table \ref{tab:semantic-scale}.
The metric \textit{relationship with super feature} can be assessed as one of five categories: sub-feature, sibling feature, super feature, identical feature, or other. 
The other four metrics were evaluated using a semantic 5-level scale ranging from 1 (poor) to 5 (excellent). 

In the second step, three authors, each holding a master degree in computer science and having five or more years of experience in software development, independently evaluated all 1200 generated feature nodes based on the evaluation guideline (1200 = 2 approaches x 20 root features x (5 nodes at Level 1 + 25 nodes at Level 2)).

Finally, in two subsequent meetings we resolved the disagreements and reached consensus. 
Final scores and labels were determined collaboratively, addressing any discrepancies through discussion and voting if necessary. 
We had less than 10\% disagreement which is a good rate according to the content analysis literature \cite{Krippendorff:ContentAnalysisIntroduction:2018}.
The scores were finally averaged across the trees to assess the performance of each approach across different domains and root features. 
During the final two meetings, we jointly tallied the number of distinct and common features across the feature trees.

\begin{table}[]
\caption{Semantic scale for assessing the generated features.}
\small
\begin{tabular}{c | p{7cm}}
\toprule
Score & Definition \\
\midrule
\multicolumn{2}{c}{Relevance} \\
\midrule
5 & Highly relevant and a logical extension of the root feature \\
4 & Mostly relevant and logically connected to the root feature \\
3 & Moderately relevant with the root feature, but may not for the same purpose \\
2 & Somewhat relevant with the root feature, because they are in the same app category \\
1 & Not relevant to the root feature at all \\
\midrule
\multicolumn{2}{c}{Clarity} \\
\midrule
5 & Very clear and easily understandable \\
4 & Mostly clear but may have some minor syntax issues \\
3 & Somewhat clear but may have some ambiguities or too long\\
2 & Mostly unclear and somewhat difficult to understand \\
1 & Very unclear or irrelevant to the sub-feature's name \\
\midrule
\multicolumn{2}{c}{Feasibility} \\
\midrule
5 & Feasible and has known examples in existing apps \\
4 & Feasible but lacks examples from existing apps \\
3 & Probably feasible but has some uncertainties \\
2 & Probably not feasible \\
1 & Not feasible \\
\midrule
\multicolumn{2}{c}{Traceability} \\
\midrule
5 & The name and description of the sub-feature are directly based on the app description with no fabrication \\
4 & The sub-feature's name is directly based on the app description, while its description is mostly based on the app description with minor fabrication \\
3 & Somewhat based on the app description but includes some fabrication \\
2 & Mostly fabricated with little relation to the app description \\
1 & Completely fabricated and not found in the app description \\
\bottomrule
\end{tabular}
\label{tab:semantic-scale}
\end{table}

\begin{table*}[!ht]
\caption{Evaluation results for the quality of generated features (L1: level 1, L2: level 2, Avg: weighted average of L1 and L2).}
\small
\begin{tabular}{l | lll | lll | lll | lll}
 & \multicolumn{6}{c|}{Existing Features} & \multicolumn{6}{c}{Novel Features} \\ \cline{2-13}
 & \multicolumn{3}{c|}{AppStore} & \multicolumn{3}{c|}{LLM} & \multicolumn{3}{c|}{AppStore} & \multicolumn{3}{c}{LLM} \\
 & L1 & L2 & Avg & L1 & L2 & Avg & L1 & L2 & Avg & L1 & L2 & Avg \\
\hline
Relevance & 5.0 & 4.94 & 4.95 & 5.0 & 4.97 & 4.97 & 3.84 & 3.91 & 3.90 & 4.96 & 4.95 & 4.95 \\
Clarity & 4.96 & 4.85 & 4.87 & 5.0 & 5.0 & 5.0 & 4.94 & 4.89 & 4.90 & 5.0 & 5.0 & 5.0 \\
Feasibility & - & - & - & 5.0 & 4.97 & 4.97 & - & - & - & 4.66 & 4.70 & 4.69 \\
Traceability & 4.91 & 4.99 & 4.97 & - & - & - & 4.98 & 4.95 & 4.96 & - & - & -
\end{tabular}
\label{tab:likert-scale-result}
\end{table*}

\begin{table*}[!ht]
\caption{Evaluation results for the relationships of generated features with their super features (L1: level 1, L2: level 2, Sum: sum of L1 and L2).}
\small
\begin{tabular}{l | lll | lll | lll | lll}
 & \multicolumn{6}{c|}{Existing Features} & \multicolumn{6}{c}{Novel Features} \\ \cline{2-13}
 & \multicolumn{3}{c|}{AppStore} & \multicolumn{3}{c|}{LLM} & \multicolumn{3}{c|}{AppStore} & \multicolumn{3}{c}{LLM} \\
 & L1 & L2 & Sum & L1 & L2 & Sum & L1 & L2 & Sum & L1 & L2 & Sum \\
\hline
Sub & 48 & 197 & 245 & 50 & 250 & 300 & 42 & 163 & 205 & 50 & 250 & 300 \\
Sibling & 0 & 23 & 23 & 0 & 0 & 0 & 0 & 24 & 24 & 0 & 0 & 0 \\
Parent & 0 & 7 & 7 & 0 & 0 & 0 & 0 & 9 & 9 & 0 & 0 & 0 \\
Identical & 2 & 14 & 16 & 0 & 0 & 0 & 1 & 12 & 13 & 0 & 0 & 0 \\
Other & 0 & 9 & 9 & 0 & 0 & 0 & 7 & 42 & 49 & 0 & 0 & 0 \\
\hline
Total & 50 & 250 & 300 & 50 & 250 & 300 & 50 & 250 & 300 & 50 & 250 & 300 \\
\end{tabular}
\label{tab:classification-result}
\end{table*}

\section{Evaluation Results} \label{sec:results}

In the following, we analyze the quality of 40 feature trees and their 1,200 recommended features obtained by refining existing and novel features with LLM-Inspiration and AppStore-Inspiration.

\subsection{Relevance}
\subsubsection{LLM-Inspiration}
As shown on Table \ref{tab:likert-scale-result}, the LLM-Inspiration achieved a high relevance score of 4.95 when refine both existing and novel features, underscoring the remarkable capability of LLM in feature recommendation. 
For instance, the feature ``Laugh Evaluation'' is described as ``continually tracks the laughs of a user to count its quantity and assesses its authenticity, emotional context, and overall impact on social interactions''.
The sub-features recommended for the root feature include ``Laugh Detection'', ``Authenticity Assessment'', ``Emotional Context Analysis'', ``Social Interaction Impact'', and ``Laugh Quantity Tracking'', all of which are highly pertinent to the root feature.

\subsubsection{AppStore-Inspiration}
The AppStore-Inspiration also demonstrates high relevance when refining existing features. 
However, for novel features, the AppStore-Inspiration yielded a relevance score of 3.90, significantly lower than the 4.97 score for existing features (Wilcoxon–Mann–Whitney test $p\leq0.00$). 
This difference can be attributed to the lack of corresponding relevant features in Google Play. 
When refining existing features with the AppStore-Inspiration, it can easily identify relevant descriptions from our app description repository and extract features from them as recommended features. 
In contrast, if a feature is not present in the app description repository, the AppStore-Inspiration will retrieve descriptions that do not fully align with the queried feature. 
For instance, when refining the root feature ``Laugh Evaluation'' using the AppStore-Inspiration, the absence of directly matching apps led to the retrieval of descriptions related to face emotion detection, laughing sound effects, or behavioral observation apps instead.
This example underscores the inherent limitation of the AppStore-Inspiration in supporting the elicitation of novel features.
Although our vector database contains approximately 589k app descriptions, it needs less than 2GB of storage.
However, most LLMs have been pre-trained on corpora exceeding 1TB, including content from books, Wikipedia, news articles, and more \cite{Zhao:SurveyLargeLanguage:2023}.
This represents a much larger knowledge base than the app stores.

\subsection{Relationship with Super Feature}
\subsubsection{LLM-Inspiration}
As shown on the Table \ref{tab:classification-result}, all sub-features recommended by LLM are logically "sub" of their super features. 
Although all sub-features recommended by the LLM-Inspiration are highly relevant to their corresponding super features, we noticed a behavioral difference based on the style of the feature description. 
When the description of the feature to be refined enumerates a list of functions, such as ``Search, compare, and book flights from various airlines with real-time pricing and availability'', the recommended sub-features may be extracted from this description. 
These sub-features include ``Flight Search'', ``Real-Time Pricing'', ``Flight Comparison'', ``Booking Management'', and ``Booking Confirmation and Notifications''.

Contrasting cases are when feature descriptions does not include enumerations as for the ``Random Chat'' feature, where the description states: ``Connect with new people globally or locally using the random chat app, where each launch introduces the user to a fresh virtual pen pal''. 
The recommended sub-features for this case includes ``Global and Local Matching'', ``User Profiles'', ``Chat Interface'', ``Safety and Moderation'', and ``Random Match Algorithm'', which are not extracted from the root feature description.

\subsubsection{AppStore-Inspiration}
When examining the AppStore-Inspi\-ration results, it becomes evident that the relationships between recommended sub-features and their corresponding super-features are not as robust as with the LLM-Inspiration. 
Although the relevance of the recommended features obtained through the AppStore-Inspiration is generally high for existing features, a discrepancy remains: only 245 out of 300 recommended features are actually "sub" of their respective super-features. 
This issue is even more prevalent with novel features, where only 205 out of 300 recommended features are actually "sub features".

Additionally, there is a noticeable variation across the different hierarchical levels of features.
Specifically, features at L1, which are direct sub-features of the root, have a higher probability  to maintain an actual "sub" relationship with their super-features compared to features at L2. 
This can be explained with two main factors:

\textit{Granularity Difference Between Root Features and L1 Features}:
Root features are typically high-level functionalities such as ``Mental Health Therapy'', ``Travel Planner'', and ``Voice Translation''.
These features often represent the main functions of an app. 
In such cases, most features described in the app description are likely sub-features of the high-level feature.
L1 features, such as ``Mini-Therapy'', ``Location-based Soundscapes'', and ``Language Selection'' are more specific and detailed making it challenging to find app descriptions that entirely match them. 
Features described in the app descriptions may not always be the sub-feature of the L1 feature.

\textit{Lack of Detail in App Descriptions}:
Another factor is the insufficient detail provided in app descriptions regarding low-level features. 
App descriptions often provide a general overview of interesting features rather than a comprehensive breakdown of all features.
This lack of detailed information complicates the extraction of sub-features at a lower level, as these specific details are often omitted from the app descriptions.

\subsection{Clarity}
\subsubsection{LLM-Inspiration}
Table \ref{tab:likert-scale-result} shows that the features obtained through the LLM-Inspiration are consistently very clear.
We found that both the names and descriptions of the features recommended by the LLM are always succinct and easy to understand. 
This is unsurprising given GPT-4's strong language generation capacity.

\subsubsection{AppStore-Inspiration}
The clarity of the features obtained through the AppStore-Inspiration is only slightly inferior to those derived from the LLM-Inspiration. 
AppStore-Inspiration generate feature description by rephrasing the sentences from app description.
Occasionally, it extracts an uninformative phrase from the app description to serve as the feature description. 
For instance, the feature description for ``Easy space reservation'' is simply ``Easy space reservation'', which lacks detail.

\subsection{Feasibility}
\subsubsection{LLM-Inspiration}
For the LLM-Inspiration, most recommended sub-features for the existing root features are feasible. 
However, when refining novel root features, the LLM-Inspiration sometimes recommends infeasible features. 
The infeasibility can be attributed to two primary factors:

\begin{itemize}
\item \textit{Technological Limitations}:
Certain features are technologically infeasible.
For instance, the recommended feature, ``Thought Interpretation Algorithm'', is described as ``utilizing advanced AI and machine learning algorithms to analyze brainwave data and interpret the user's thoughts''. 
The feasibility of this feature is rated as low due to the immature state of brainwave translation technology.
\item \textit{Permission Constraints}: 
Certain features are deemed infeasible due to potential violations of user permissions or legal regulations.
For instance, the recommended feature ``Offline Interaction Logging'' involves the offline monitoring of user interactions (such as face-to-face conversations and phone calls) raising serious privacy and legal concerns.
\end{itemize}

\subsubsection{AppStore-Inspiration}
The feasibility of the features recommended with AppStore-Inspiration is not evaluated, as they are already successfully implemented by existing apps.

\subsection{Traceability}
\subsubsection{LLM-Inspiration}
Traceability is not evaluated for features recommended with LLM-Inspiration.
This limitation arises from the inherent difficulty in distinguishing whether a feature recommended is an original creation of the model or if it has been extracted from its extensive training corpus.

\subsubsection{AppStore-Inspiration}
In the AppStore-Inspiration, traceability is generally excellent. 
Most of the recommended features can be directly traced back to their respective app descriptions.
Only a small number of recommended features cannot be linked to the source sentences from the app description. 
This indicates the capability of GPT-4 to effectively extract features from app descriptions.

An interesting observation is that at least ten apps were no longer available on Google Play at the time of our evaluation, which occurred two months after we collected the app descriptions. 
This did not impact our evaluation of traceability, as we saved the app descriptions in our repository.

\subsection{Redundancy}

\subsubsection{LLM-Inspiration}
Table \ref{tab:distinct-features} presents the number of distinct features.
As there are 30 recommended features in a feature tree, this finding indicates minimal redundancy within the features. 
This phenomenon can be attributed to the impressive reasoning capabilities of GPT-4, which enables it to generate sub-features that precisely align with their respective super-feature descriptions. 
Consequently, the recommended features remains distinct, effectively reducing redundancy and enhancing the granularity of the feature tree.

\begin{table}[!h]
\caption{Average number of distinct features of a feature tree.}
\small
\begin{tabular}{l | cc | cc}
& \multicolumn{2}{c|}{Existing Root} & \multicolumn{2}{c}{Novel Root} \\
& AppStore & LLM & AppStore & LLM \\
\hline
\# of Distinct Features & 22.3 & 29.6 & 21.9 & 30 \\
\# of Distinct Relevant Features & 21.6 & 29.3 & 14.5 & 30 \\
\end{tabular}
\label{tab:distinct-features}
\end{table}

\subsubsection{AppStore-Inspiration}
In contrast, the AppStore-Inspiration exhibits a more severe redundancy problem. 
For instance, in the tree derived from the ``Anti-Smartphone Addiction'' root feature, the ``Daily App Limit'' feature appears multiple times. 
Specifically, it is present once at level 1 and three times at level 2 as a sub-feature under ``Customizable Time Restrictions'', ``Screen Time Tracking'', and ``Time Blocking''.
We hypothesize that this redundancy stems from the limited variety of features described within the app descriptions, which forces the approach to reuse extracted app features when refining different features.

\subsection{Common and Different Features}
\begin{figure}[!htb]
    \centering
    \includegraphics[width=0.45\textwidth]{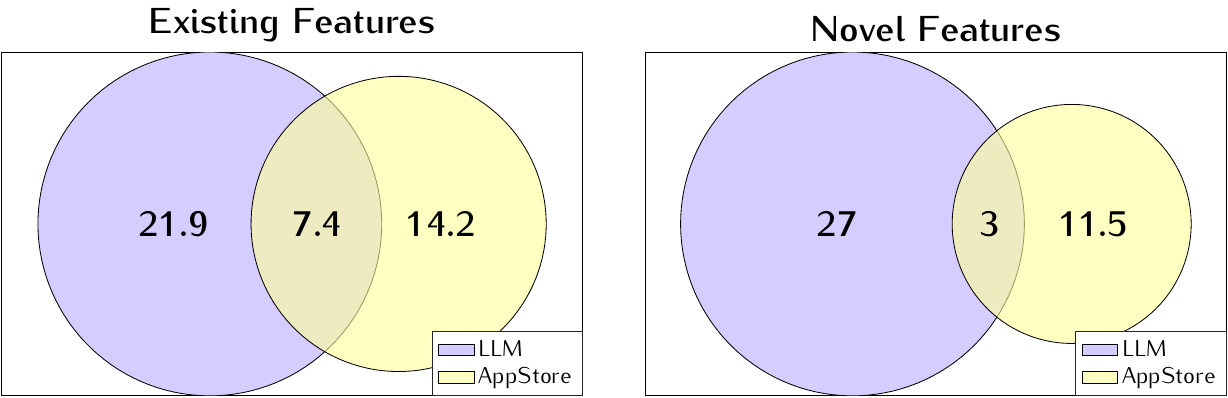}\\
    \caption{Venn diagrams showing the average number of common and different features (distinct and relevant) generated by LLM-Inspiration and AppStore-Inspiration.}
    \Description{}
    \label{fig:features-intersection}
\end{figure}

Figure \ref{fig:features-intersection} illustrates the average number of common and different (distinct and relevant) features in the feature trees obtained by LLM-Inspiration and AppStore-Inspiration.
Irrelevant features are not included in this count.
The figure shows that the intersection is small: only 7.4 features when refining existing feature.
When refining novel features, the common feature count is only 3.

The difference set between the two approaches is even larger than their intersection set.
For existing features, the primary reason for this substantial difference is the granularity of the features. 
Features of different granularity do have an overlap. 
However, they were not considered as the same feature during our evaluation.
While for novel features, the main reason is the variety of solutions, for example, when refining the ``Though Reading'' feature, LLM-Inspiration tends to do it by ``Brainwave detection'', while AppStore-Inspiration proposes ``Judge by body language''.
These two reasons explain most of the differences for both existing and novel features.
In the following, we discuss additional reasons specific to each approach.

\subsubsection{LLM-Inspiration}
We observed that the difference feature set of the LLM-Inspiration is mainly due to their absence in app descriptions.
This omission can be attributed to two reasons: either the basic features are deemed too trivial to mention, or the features are not implemented in other apps.

\textit{Trivial Features}:
The LLM-Inspiration refines a feature according to its description, producing sub-features that precisely align with the parent. 
However, these generated sub-features may be either too fundamental or not sufficiently engaging, prompting app vendors to exclude them from app descriptions. 
For example, a feature tree generated from the root ``Supermarket Checkout'' using the LLM-Inspiration might include ``Scan History'', ``Error Handling'', ``View Cart'', and ``Remove Item'' as sub-features. 
While essential, these sub-features are relatively basic and may not be considered by vendors noteworthy to be highlighted in the app descriptions, that do not aim to provide a complete feature overview.

\textit{Innovative Features}:
Some features generated by the LLM-Inspi\-ration do not exist in current apps, such as ``AR Historical Visualization'', ``Mood Detection via Device Sensors'', and ``Thought Interpretation Algorithm''. 
This phenomenon occurs particularly when refining novel features, which have not yet been widely adopted in existing apps lacking corresponding app descriptions.

\subsubsection{AppStore-Inspiration}
Unlike the LLM-Inspiration, which refines a feature based on its description, the AppStore-Inspiration tends to recommend features with some degree of relevance but not entirely aligned with its super feature's description.
This results in the following two types of features:

\textit{Features with Additional Information}:
This often occurs with features at level 2. 
Because the description of a level 1 feature may contain additional information, which may suggest more features when refining.
For example, the description of the L1 feature ``Personalized Style Recommendations'' includes the phrase ``app's AI algorithms analyze your fashion preferences'', which is not mentioned in its super feature ``Virtual Fashion Assistant''.
Consequently, when refining the ``Personalized Style Recommendations'' feature, the AppStore-Inspiration recommends ``AI Personal Stylist''.

\textit{Cross Domain Features}:
When exploring feature inspiration with novel features, even in the absence of an app that precisely matches the description of these new features, AppStore-Inspiration may suggest cross-domain features. 
For instance, in refining the root feature ``Synesthetic Sensory Augmentation'', the AppStore-Inspiration identified numerous relevant features from meditation apps, like ``Drawing sound'' and ``Discover Synesthesia''.
They may not exactly fit the purpose of the root feature ``Synesthetic Sensory Augmentation'', which is intended for interaction with digital content, but they do provide valuable inspiration.


\section{Discussion}
\label{sec:disc}
\subsection{LLMs vs. App Stores for Feature Inspiration}
While both approaches seem to be able to recommend relevant sub-features in most cases, upon comparing LLM-Inspiration and AppStore-Inspiration, we found that LLM-Inspiration to be more powerful.
The sub-features recommended by LLM-Inspiration are highly relevant to their corresponding super features even for novel app scopes. 
Moreover, they are consistently logical extensions of their super features. 
Most recommended sub-features are feasible, even when refining novel features. 
But some seem imaginary, which suggests the importance of a human-analyst in the elicitation loop \cite{Andersen:DesignPatternsMachine:2024, Wei:AIInspiredUIDesign:2024}.  
We think that LLM-Inspiration can support or partially replace humans in the feature refinement task particularly for preliminary iterations.

It is important to note that the LLM-Inspiration is likely sensitive to the description of the super feature, suggesting that practitioners may need to experiment with and adjust the description to achieve optimal results.
We did not study the impact of the feature description (quality) on the generated trees \cite{Montgomery:EmpiricalResearchRequirements:2022}. 
It is, e.g., likely that short/long or redundant/varying descriptions, as well as descriptions pointing to a solution or a technology will impact the recommended sub-features. 
In the future, researchers may investigate this impact in developer studies and benchmarking experiments. 

The sub-features recommended by AppStore-Inspiration exhibit high relevance when refining high-level and existing features too.
However, when it comes to low-level or novel features---a more advance brainstorming and reasoning task---the recommended features may not always logically align as "sub" of their respective super features. 
These sub-features require filtering and editing before reuse. 
Despite this issue with relevance, a tool supporting feature elicitation may still recommend interesting cross-domain features. 
One significant advantage of AppStore-Inspiration is that each recommended feature is linked to its source app. 
This linkage allows practitioners to explore the source app for implementation details and user feedback on the features \cite{Johann:SAFESimpleApproach:2017,Haering:AutomaticallyMatchingBug:2021}.

\subsection{Tool Support}
\begin{figure}[!htb]
\centerline{\includegraphics[width=0.5\textwidth]{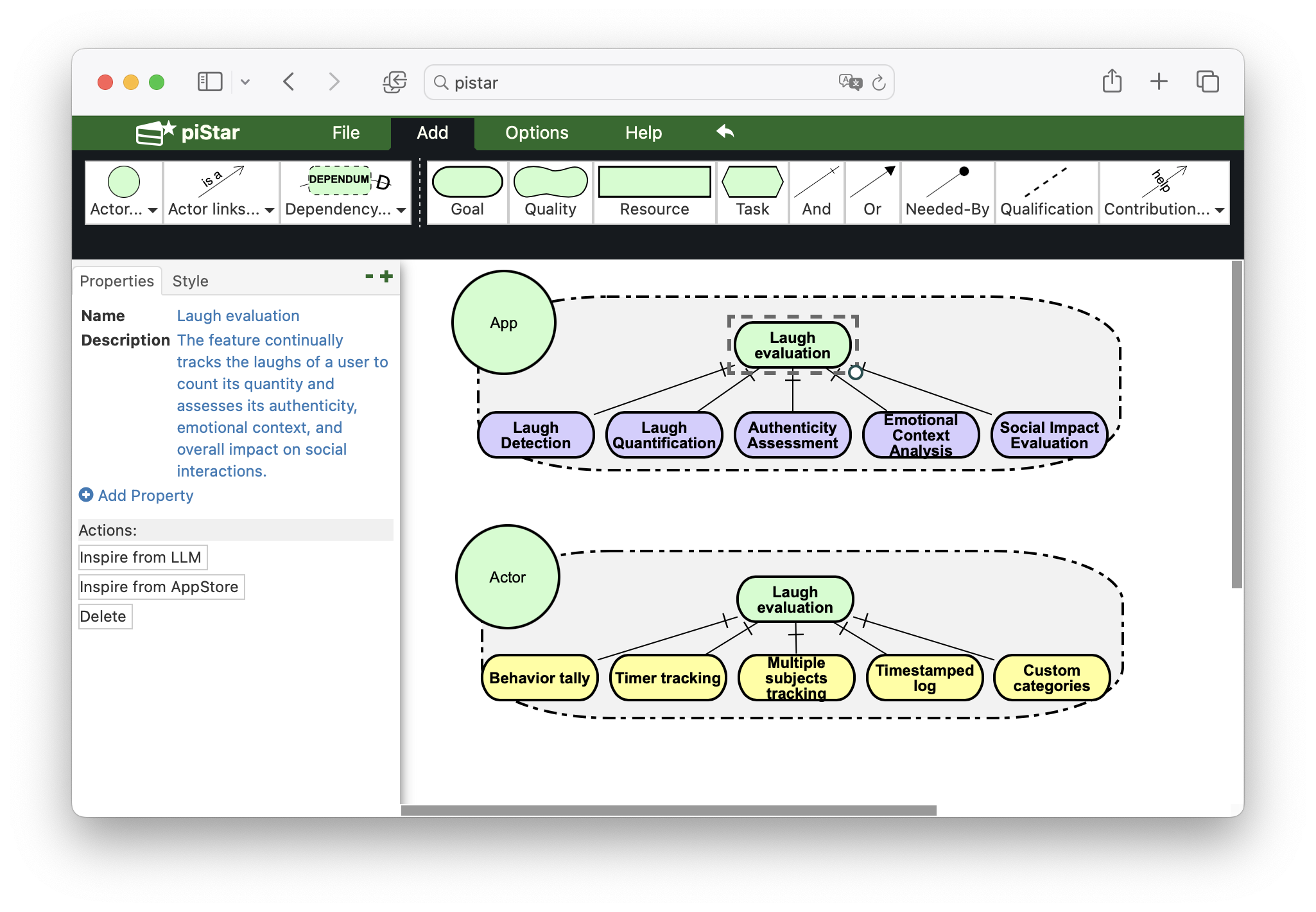}}
\caption{PiStar integrating new buttons (left sidebar) for inspiration from LLM or AppStore.
By clicking ``Inspire from LLM'' or ``Inspire from AppStore'' for a feature, corresponding sub-features will be generated automatically.
Sub-features generated with LLM are in purple, those from AppStore are in yellow.}
\Description{}
\label{fig:pistar}
\end{figure}

We have implemented our LLM-Inspiration and AppStore-Inspiration within piStar \cite{Pimentel:PiStarToolPluggable:2018}, a goal modeling tool, to facilitate the adoption, as shown on the Figure \ref{fig:pistar}.
Goal models, such as KAOS \cite{Respect-IT:KAOSTutotial:2007} and i* \cite{Yu:SocialModeling:2009}, are well-known in requirements engineering.
The goal model is constructed by asking ``why'' and ``how'' questions starting from a root node.
The ``how'' question will derive sub-goals, which is very similar to the feature refinement process discussed in this paper.

We preliminary tested this tool with three experts in software engineering (professors not contributing to this work). 
The experts were firstly asked to manually refine the feature ``laugh evaluation'' into a two-levels feature tree without the assistance of LLM or AppStore-inspiration. 
Subsequently, they generated sub-features using LLM and AppStore-inspiration. 
We then asked them for their feedback about the results from both approaches and whether they think this tool is helpful.
Their feedback indicates that both approaches can save time and provide useful inspirations. 
It seems that the LLM-inspiration was found more interesting, although the recommended sub-features may not be perfect, experts will likely complete or correct them.
It seems that it is easier to refine and edit these features than to create them from scratch.

In real-world scenarios, we believe that automated feature elicitation should be an iterative process that actively involves developers and analysts at every stage \cite{Wei:AIInspiredUIDesign:2024}. 
Multiple iterations are necessary, allowing analysts to review and refine feature names and descriptions as needed. 
Moreover, with LLM-based approaches, feedback can be provided directly to the model. 
For example, an analyst might specify, ``This sub-feature is not relevant because... Please generate 10 alternative sub-features''. 
This iterative feedback loop ensures more accurate and meaningful feature development.

Obviously, the two feature recommendation approaches studied in this paper can be easily integrated into other software engineering tools as well \cite{Maalej:TaskFirstContextFirstTool:2009}. 
This can be for instance, an extension of an issue tracker (as Jira, GitHub Issues or Trac) to assist practitioners create and prioritize epics and feature requests. 
Also collaboration and brainstorming tools such as Miro or Conceptboard are suitable to include LLM- and AppStore- based inspirations for sub-features.

\subsection{Threats to Validity}
This section discusses potential threats to the validity of our study.

\paragraph{Limitated Number of App Descriptions}
There are more than 2.43 million apps on Google Play. 
However, we have been only able to collect 849 k apps from this largest app store.
This limitation could potentially affect the results obtained by the AppStore-Inspiration.
To mitigate this issue, we ensured that all evaluated existing features are actually present in our dataset, and any novel features included in our study do not currently exist on Google Play.
These steps helped to validate our results despite the smaller set of apps, ensuring that the conclusions drawn remain robust and reliable.

\paragraph{Selection of Root Features}
In comparison to the nearly infinite number of app features, the 20 root features used in our evaluation may seem limited. 
This limitation arises from the considerable manual effort required, as evaluating each root feature required the manual assessment of 60 sub-features from both approaches by three authors. 
This necessitates a balance between the feasibility of labeling (i.e. needed effort) and the sample size.
To maximize the generalizability of our study, we included 10 novel features from 10 different app categories and 10 existing features from 10 different app categories. 
Additionally, we evaluated not only the 20 root features but also all the 200 sub-features derived from them using both approaches. 
Therefore we believe that our evaluation covers a fairly broad and representative scope.
Certainly, replicating the study with other types of features and from other domains will further strengthen the generalizabiltiy of the results. 

\paragraph{Subjectivity in Manual Evaluation}
As for every manual research task, subjectivity and potential observer bias might lead to variations in how different evaluators interpret and assess the generated features. 
To mitigate this potential threat, we 
(1) created an evaluation guide including a well defined semantic scale to detail the definition of each score with examples,
(2) evaluated each generated features independently by three evaluators, 
and (3) reviewed the final scores and labels through discussion and consensus. 
Overall, the evaluators, who possess five or more years of software development experience, did not think that the evaluation of generated features was a complex task. 
This is also reflected in the fairly high rate of achieved agreements. 
Nonetheless, it is important to focus on the comparative trends when interpreting our results rather than the exact scores.

\paragraph{Maturity of Implementation}
It is inappropriate to compare the speed of a sports car with that of a steam train and conclude that the car is faster. 
Similarly, it is hard to compare the performance of a prototypical implementation of the AppStore-based approach with a well implemented LLM-based approach and assert that the LLM is superior to AppStore. 
To mitigate this issue, we also relied on GPT-4 in the feature extraction and feature selection stages of the AppStore-Inspiration, and ensured that the system prompts for both LLM and AppStore are very similar. 
It is also important to note that this work did not evaluate each individual step of the AppStore-Inspiration (which might be considered in future work). 
We are thus unable to discuss the impact of each steps. 
However, we still think that each step is necessary for the final feature recommendation as our evaluation of the ``relevance'' and ``traceability'' overall suggest.
Moreover, for both approaches, we did not use advanced prompt engineering techniques such chain-of-thought prompting which might impact the results too.

\section{Related work}
\label{sec:relwork}
\subsection{App Store Mining for Requirements}
As shown by Ferrari and Spolitini  \cite{Ferrari:StrategiesBenefitsChallenges:2023}, app stores serve as an important source for inspiring requirements elicitation.
App stores contain various data, including app descriptions, app reviews, and app images.
We summarize existing work in these areas.

\subsubsection{Mining App Descriptions}
App descriptions, composed by the application developers and vendors themselves, provide a succinct introduction to the salient features of the respective apps.
Recent studies have sought to mine these app descriptions in a variety of manners.
This includes the identification of similar apps by analyzing their respective descriptions \cite{Uddin:AppCompetitionMatters:2020,Hamednai:SimAndroEffectiveMethod:2019,Al-Subaihin:EmpiricalComparisonTextbased:2019},
and the extraction of app features from the app descriptions \cite{Johann:SAFESimpleApproach:2017}. 
The extracted features can be used to construct domain knowledge \cite{Liu:MiningDomainKnowledge:2017}.
In addition, these features serve as a basis for recommending requirements, as evidenced by several studies \cite{Jiang:RecommendingNewFeatures:2019,Liu:InformationRecommendationBased:2019,Liu:AppStoreMining:2019,Liu:MethodAcquireCrossdomain:2022}.
Our work aims at comparing app mining approaches to recent general purpose LLMs.

\subsubsection{Mining App Reviews}
Reviews on app stores provide valuable insights from users, for example, the feature requests or bug reports, making them a valuable resource for requirements elicitation \cite{Pagano:UserFeedbackAppstore:2013, Gomez:AppStoreCrowdsourced:2017}.
Given the vast volume of app reviews, researchers have introduced numerous techniques to enhance the efficiency of their analysis. 
These techniques encompass the automatic classifications of app reviews into predefined category such as bug reports and feature requests \cite{Maalej:AutomaticClassificationApp:2016,Stanik:ClassifyingMultilingualUser:2019,Devine:EvaluatingSoftwareUser:2023,Wei:ZeroshotBilingualApp:2023,Wei:DataDrivenRequirementsEngineering:2022}.
Additionally, these methods employ clustering algorithms to assemble app reviews based on semantic similarity \cite{Scalabrino:ListeningCrowdRelease:2019,Stanik:UnsupervisedTopicDiscovery:2021,Devine:WhatClusterSoftware:2022,Wei:ZeroshotBilingualApp:2023}, and also involve the generation of concise summaries of app reviews \cite{DiSorbo:SURFSummarizerUser:2017,Devine:WhatClusterSoftware:2022,Harkous:HarkDeepLearning:2022,Wei:ZeroshotBilingualApp:2023}.
These techniques are complementary to AppStore- and LLM-Inspiration as they bring the perspective and creativity of end users. 

\subsubsection{Mining App Introduction Images}
The app introduction images on Google Play are a gold mine for the inspiration of app design, particularly the Graphical User Interface (GUI), as they are carefully selected by app developers to represent the important features of the apps.
Recent researches mines the app introduction images and proposed GUI search engines, such as Gallery D.C. \cite{Feng:GalleryAutocreatedGUI:2022,Chen:GalleryDesignSearch:2019}, and GUing \cite{Wei:GUingMobileGUI:2024}, to facilitate the search of existing app UI designs using textual queries.
Recently, Wei et al.~discussed how LLM-Inspiration can be combined with GUI-Mining with the app designer in the loop \cite{Wei:AIInspiredUIDesign:2024}.

\subsection{LLMs for Requirements Elicitation}
Particularly since the release of ChatGPT, numerous studies have investigated the capacity of large language models (LLMs) for facilitating requirements elicitation. 
For instance, Ronanki et al. \cite{Ronanki:InvestigatingChatGPTPotential:2023} examined the potential of ChatGPT in assisting the requirements elicitation process, concluding that ChatGPT-generated requirements are notably more abstract, atomic, consistent, correct, and understandable compared to those formulated by human experts. 
Gorer et al. \cite{Gorer:GeneratingRequirementsElicitation:2023} used LLMs for generating requirements elicitation interview scripts, demonstrating the model's efficacy in enhancing the quality of these scripts.
Cabrero-Daniel et al. \cite{Cabrero-Daniel:ExploringHumanAICollaboration:2024} investigated the utilization of GPT-4 as assistants in agile software development meetings.
Additionally, Marczak-Czajka et al. \cite{Marczak-Czajka:UsingChatGPTGenerate:2023} applied ChatGPT to generate human-value user stories, thus providing inspiration for new requirements. 
In a similar vein, Zhang et al. \cite{Zhang:LLMbasedAgentsAutomating:2024} utilized GPT models to evaluate and refine the quality of user stories. 
Furthermore, Ataei et al. \cite{Ataei:ElicitronLLMAgentBased:2024} developed multiple agents based on GPT-4, which facilitated the exploration of a broader range of user needs and unanticipated use cases.
Recent work by Chen et al. \cite{Chen:UseGPT4Creating:2023} and Nakagawa et al. \cite{Nakagawa:MAPEKLoopbasedGoal:2023} has underscored the potential of generating goal models from given contexts using LLMs. 
These advancements collectively highlight the growing capability of LLMs in various aspects of requirements elicitation and analysis.

\section{Conclusion}
\label{sec:conclusion}
Since the emergence of app stores and of mining techniques for app data, app stores-inspired requirements elicitation is getting more and more popular in industry \cite{Martens:ReleaseEarlyRelease:2019}. 
The main advantage is the large knowledge about apps, what they offer and how users react to them.
With the advent of LLMs, their impressive capabilities reveal the potential of (LLM)-inspired requirements elicitation as well. 
However, the current literature offers limited insights into a comparison between both approaches.
In this work, we implemented LLM-Inspiration and AppStore-Inspiration for the task of feature refinement. 
We performed manual evaluation on 1,200 sub-features obtained through both approaches.
Our findings indicate that the LLM-based approach recommends highly relevant sub-features, potentially even partially replacing human effort in feature refinement. 
An AppStore-based approach seems better at recommending cross-domain apps and validate a feature and its feasibility by exploring its source app. 
In practice, a careful combination of both approaches will likely lead to the best results. 
In future works, we plan to: 
(1) integrate the app reviews into our AppStore-based approach to better inform the importance of the suggested features and 
(2) survey practitioners to evaluate how these two approaches are used and should be used in their workflows.

\balance
\bibliographystyle{ACM-Reference-Format}
\bibliography{ref}

\end{document}